# Hierarchical correlations in models of galaxy clustering

E. Gaztañaga[1] & C.M. Baugh[1,2]

[1] Department of Physics, Keble Road, Oxford OX1 3RH
[2] Physics Dept., University of Durham, South Road, Durham DH1 3LE

**ABSTRACT**

We present a comparison of the predictions of perturbation theory for the hierarchical J-order correlation amplitudes $S_J = \overline{\xi}_J / \overline{\xi}_2^{J-1}$, $J = 3 - 10$, with the results of large numerical simulations of gravitational clustering. We consider two different initial power spectra of density fluctuations, one being flatter than the other. The clustering amplitudes $S_J$ measured in these models are different at all scales. In each case, the perturbation theory predictions give an excellent agreement with the simulations on scales for which the variance is approximately linear, i.e. for scales on which $\overline{\xi}_2 \lesssim 1$. For cells of radius $R \gtrsim 30 \, h^{-1} \, \mathrm{Mpc}$, the sampling variance arising from the finite size of our simulations dominates the results. We also find that $S_J$ are roughly independent of time, $\Omega$, or $\lambda$, as expected. On comparing with the observations, one can use these results to discriminate between models for structure formation.

## 1 INTRODUCTION

The statistical properties of a density field can be characterised in a simple and efficient way by the J-order connected moments (or volume-averaged J-point correlation functions) $\overline{\xi}_J(R) \equiv < \delta_R^J >_c$ of density fluctuations $\delta_R$ averaged in spherical volumes of radius $R$ (for definitions of the notation used in this Letter, see Gaztañaga 1994).

Gravitational evolution from an initially Gaussian density field generates higher order correlations, $\overline{\xi}_J$ with $J > 2$, which are given by the so called 'hierarchical' constants: $S_J = \overline{\xi}_J / \overline{\xi}_2^{J-1}$. The values of $S_J$ for density fluctuations in spherical cells have been estimated using perturbation theory (PT) in the limit $\overline{\xi}_2 \rightarrow 0$ (e.g. Juskiewicz et al. 1993, Bernardeau 1994b). In this limit, the $S_J$ are independent of time and they are a function only of the shape of the initial variance $\overline{\xi}_2$. Baugh, Gaztañaga & Efstathiou (1994, hereafter BGE94) have performed a detailed systematic comparison of the predictions of PT with the results of N-body simulations, for models with an initial two-point correlation, $\overline{\xi}_2$ given by the standard cold dark matter scenario (SCDM), i.e. for a universe with $\Omega = 1$ and $h = 0.5$, specified by $\Gamma = \Omega h = 0.5$ ($h$ denotes Hubble's constant in units of $100 \, h \, \mathrm{kms^{-1} Mpc^{-1}}$). Using simulations with different numbers of particles and different computational box sizes, BGE94 confirm the conclusions of previous studies (e.g. Juskiewicz et al. 1993, Kofman & Bernardeau 1994, Bernardeau 1994a) and extend them to higher orders. The main result of these analyses is that the PT predictions for the form of the $S_J$ in the SCDM model agree with the values measured from the simulations for scales where $\overline{\xi}_2 \lesssim 1$. In this regime, according to PT, the values of $S_J$ should be independent of time, also in agreement with what is found in simulations (BGE94). These results hold for other shapes of $\overline{\xi}_2$ and the values of $S_J$ are almost independent of the values of the density parameter $\Omega$ or the cosmological constant $\lambda$ (Bouchet et al. 1992, Bernardeau 1994a).

These results are important because they indicate that PT can provide us with predictions for the properties of a density field undergoing gravitational evolution which are independent of $\Omega$, $\lambda$ or time. In contrast the increase in the amplitude of the 2-point function, $\overline{\xi}_2$, due to gravitational evolution does depend upon time, $\Omega$, $\lambda$ and the biasing of the light distribution with respect to the mass. Hence there are several possible factors that could contribute to the 2-point correlations, so that a comparison with the clustering in galaxy catalogues would not necessarily distinguish between them. In particular, it would be difficult to argue that we have conclusive evidence that fluctuations grow by gravitational evolution by just using the observations of $\overline{\xi}_2$. Higher order moments $S_J$ can help to remove some of this ambiguity. They also provide us with a powerful tool to study how galaxy fluctuations trace matter fluctuations (see Gaztañaga & Frieman 1994).

In this letter, we compare the PT predictions with estimations made from large N-body simulations for two models which have different correlations $\overline{\xi}_J$ at all scales. For comparison, we have also made artificial point distributions with an identical size box, number of particles and values of $\overline{\xi}_2$, and roughly the same number of clusters and mass function in clusters. We use this control sample to test if, in realistic but non gravitational models with clumps, there is also correlation between the values of $S_J$ and the shape of $\overline{\xi}_2$.

## 2 SIMULATIONS

### 2.1 N-body models

The initial linear theory power spectrum of the density fluctuations used in the simulations is that of Bond & Efstathiou

(1984) which is a function of $\Gamma \equiv \Omega h$. This form for the power spectrum applies for scale-invariant CDM universes that have low baryon densities, $\Omega_B \sim 0.03$. We have simulated two models that we label $\Gamma = 0.5$ and $\Gamma = 0.2$. The $\Gamma = 0.5$ model corresponds to the SCDM model, $\Omega = 1$ and $h = 0.5$, while the $\Gamma = 0.2$ model is a spatially flat low density CDM variant, with a cosmological constant $\lambda = 0.8$, $h = 1$ and $\Omega = 0.2$ at the present epoch. Although these are CDM motivated models, what is relevant here is that the shapes they predict for the two-point statistic, $P(k)$ or $\overline{\xi}_2$, are different. The results we obtain would also apply for other models that produce similar shapes for $\overline{\xi}_2$.

We ran ensembles of 5 simulations for each model, using the same random phases, in a box of size $L_B = 378\,h^{-1}$ Mpc containing $N = 126^3$ particles. The simulations are evolved using the $P^3M$ code of Efstathiou *et al.* 1985.

The volume averaged correlations $\overline{\xi}_J$ are estimated from moments of counts-in-cells as in BGE94, averaging the results over the simulations in each ensemble and using the dispersion between members of the ensemble to estimate the sampling error. Different stages in the evolution of the simulations are labelled in terms of the linear PT variance in spheres of radius $8\,h^{-1}$ Mpc, denoted $\sigma_8$, calculated from the initial power spectrum. After a number $a$ of expansion factors the initial linear value $\sigma_8^i$ increases to: $a\sigma_8^i$. In all our simulations the initial value of $\sigma_8$ is $\sigma_8^i \simeq 0.10$ and the final value is $\sigma_8^f \simeq 1.0$ so that they are evolved over 10 expansion factors. Nevertheless, the time resolution for $S_J$ is effectively about 3 expansion factors as higher order correlations are initially affected by a transient arising from the the original displacements of the particles from a grid made using the Zeldovich approximation (see BGE94).

## 2.2    Clumps model

For comparison we also make a set of "mock" point distributions based on a combination of the clustering models discussed in Peebles (1980, §61). Clusters are made of spherical clumps which are assumed to have $N_g$ galaxies (or matter points) with a number density varying as a power of distance from the clump center. To introduce a hierarchical structure, clumps are placed at each step of a Rayleigh-Lévy random walk. Starting from a random position in a periodic box, the next clump in the sequence is placed in a randomly chosen direction at a distance drawn from another power law distribution. Each clump has a different number of galaxies with a "mass function" which roughly imitates that of the simulations.

Of course, this model is not a true representation of the physical process underlying galaxy formation. We have chosen it because it is convenient and seems to reproduce some of the properties of gravitational clustering. In particular, one can adjust the parameters in the model to obtain a given shape and amplitude for $\overline{\xi}_2$. We have run several of these models but shall concentrate on ensembles that reproduce closely the $\Gamma = 0.2$ shape for $\overline{\xi}_2$. We have generated two sets of 5 of these models for each ensemble corresponding to different amplitudes of $\overline{\xi}_2$ in a box of size $L_B = 378\,h^{-1}$ Mpc containing $N = 126^3$ particles. Half a slice from the central part of one of these simulations is shown in Figure 1, where it is compared with slices in the $\Gamma = 0.2$ and $\Gamma = 0.5$ models.

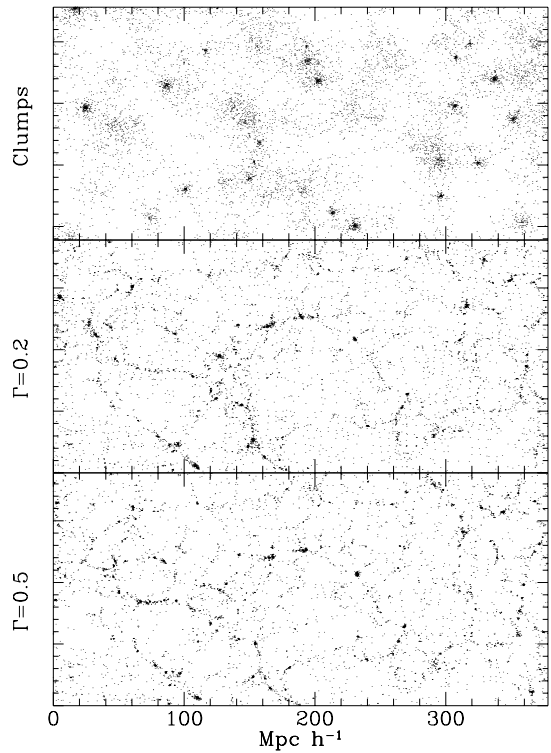

**Figure 1.** Slices of $3 \times 189 \times 378\,h^{-1}$ Mpc, in the center of the simulated box in the clumps, $\Gamma = 0.2$ and $\Gamma = 0.5$ models.

## 3    RESULTS

### 3.1    Clustering in N-body models

In linear PT the evolution of the density field is completely specified by the 2-point correlation which, for $\Omega = 1$, grows with the scale factor $a$ as $\overline{\xi}_2(R, t) = a^2 \overline{\xi}_2(R, t_0)$, where $R$ is the comoving scale and $t_0$ is some initial time. The linear growth is illustrated in Figure 2, which shows the values of $\overline{\xi}_2$ for two expansion factors in the simulations. The errors are always smaller than the size of the symbols. The solid (dashed) lines show the linear PT prediction $\overline{\xi}_2^{(L)}$ for the $\Gamma = 0.5$ ($\Gamma = 0.2$) model, obtained from a numerical integration of the initial power spectrum,

$$\overline{\xi}_2 = \frac{V}{2\pi^2} \int_0^\infty \mathrm{d}k\, k^2 P(k) W^2(kR), \qquad (1)$$

where $W(kR)$ is the Fourier transform of the spherical window with radius $R$. Each line is normalized using the expansions factors $a$ obtained from the simulation. The most noticeable discrepancy due to nonlinearities in Figure 2 appears at scales $R \lesssim 10\,h^{-1}$ Mpc where the N-body results give larger density fluctuations than the linear PT prediction. Note however that although the $\Gamma = 0.2$ model has less power at small scales in the initial conditions it has more power than the $\Gamma = 0.5$ model at all scales for later output times.

The hierarchical scaling predicts $\overline{\xi}_J = S_J\,\overline{\xi}_2^{J-1}$ where the $S_J$ are roughly constant. Bernardeau (1994b) has estimated $S_J$ using PT for spherical fluctuations (top-hat win-

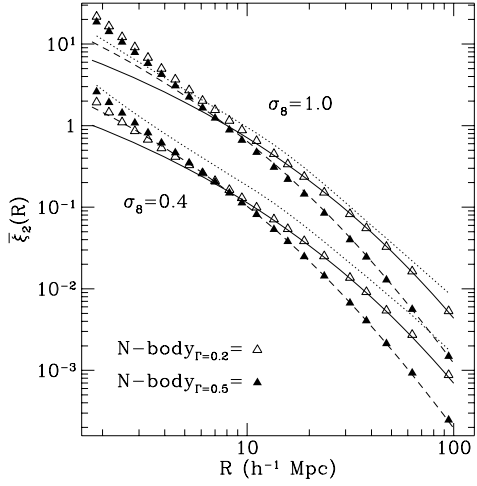

**Figure 2.** The variance $\overline{\xi}_2$ at two different epochs, $\sigma_8 = 0.40$ and $\sigma_8 = 1$, estimated from N-body simulations: models $\Gamma = 0.2$, open symbols, and $\Gamma = 0.5$, closed symbols. The solid (dashed) lines show the linear PT predictions for the $\Gamma = 0.2$ ($\Gamma = 0.5$) model. Dotted lines correspond to clumps models with $\sigma_8 \simeq 1.2$, upper line, and $\sigma_8 \simeq 0.5$, lower line.

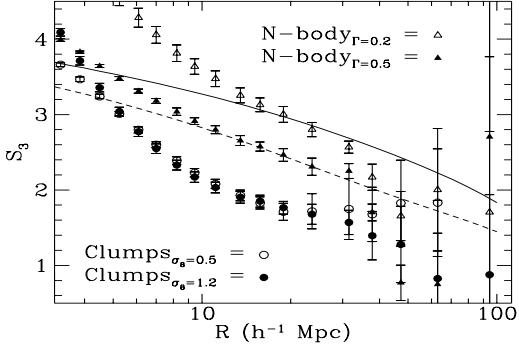

**Figure 3.** The hierarchical amplitudes $S_3 = \overline{\xi}_3/\overline{\xi}_2^2$ in the $\Gamma = 0.5$ CDM model (filled triangles) compared with the $\Gamma = 0.2$ CDM model (open triangles). The solid (dashed) lines show the linear PT predictions for the $\Gamma = 0.2$ ($\Gamma = 0.5$) model. Open (filled) circles correspond to the $\sigma_8 = 0.5$ ($\sigma_8 = 0.2$) clumps model.

dow) in terms of the derivatives of the linear $\overline{\xi}_2$:

$$\gamma_J \equiv \frac{d^J \log \overline{\xi}_2}{d \log^J R}. \qquad (2)$$

For example, for $J = 3$ and $J = 4$, Bernardeau finds:

$$
\begin{aligned}
S_3 &= \frac{34}{7} + \gamma_1 \\
S_4 &= \frac{60712}{1323} + \frac{62}{3}\gamma_1 + \frac{7}{3}\gamma_1^2 + \frac{2}{3}\gamma_2.
\end{aligned} \qquad (3)
$$

Thus, $S_J$ are not constant as the shape of $\overline{\xi}_2$, and therefore $\gamma_J$, changes with scale. For our comparison we have set high order derivatives $\gamma_J = 0$ for $J > 2$ (see BGE94). The above values correspond to $\Omega = 1$ and $\lambda = 0$, but the dependence on $\Omega$ and $\lambda$ is negligible (Bouchet *et al.* 1992, Bernardeau 1994a).

Figures 3-5 show the amplitudes $S_J$ in the $\sigma_8 = 1$ output time of the simulations, estimated from the ratios $\overline{\xi}_J/\overline{\xi}_2^{J-1}$. In all figures, we also show the predictions from PT as solid lines for the $\Gamma = 0.2$ model and dashed lines for the $\Gamma = 0.5$ model. Errors corresponding to the dispersion over 5 simulations are larger in $\Gamma = 0.2$ model. On comparing the values of $\overline{\xi}_J$ for individual simulations one finds that they do not fluctuate around the mean as a function of $R$ but can be significantly shifted at all scales with respect to the mean (see BGE94). These overall shifts are produced by density fluctuations on scales comparable to the box size which are not properly represented in each individual simulation. According to this interpretation we would expect the $\Gamma = 0.2$ model to have larger sampling errors as it has bigger density fluctuations at large scales (Figure 2).

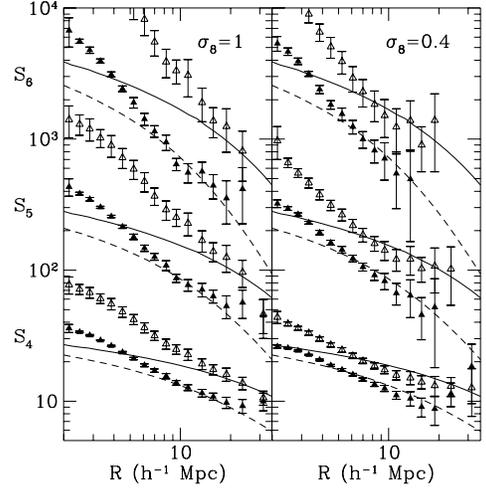

**Figure 4.** The hierarchical amplitudes $S_J = \overline{\xi}_J/\overline{\xi}_2^{J-1}$ for $J = 4-6$ in the $\Gamma = 0.5$ (filled triangles) and $\Gamma = 0.2$ (open triangles) models, at two different output times, $\sigma_8 = 1.00$ (left panel) and $\sigma_8 = 0.40$ (right panel).

In Figure 4 we also show the $\sigma_8 = 0.40$ output time for comparison. Note that the results are very similar in both output times, indicating that the values of $S_J$ do not evolve significantly with time.

### 3.2 Clustering in the Clumps model

The clumps simulations are constructed to reproduce the variance of the $\Gamma = 0.2$ N-body model. Dotted lines in Figure 2 show $\overline{\xi}_2$ for two ensembles (each with 5 simulations) with different amplitudes: $\sigma_8 \simeq 1.2$, upper line, and $\sigma_8 \simeq 0.5$, lower line. The values of $S_3$ are shown in Figure 3 as open ($\sigma_8 \simeq 1.2$) and closed ($\sigma_8 \simeq 0.5$) circles. These values of $S_3$ are significantly different to the values in the N-body models. Discrepancies are even larger for higher orders: e.g., at $R \simeq 8 \ h^{-1}$ Mpc the clumps model gives $S_7 \simeq 370$ compared with $S_7 \simeq 56000$ in the $\Gamma = 0.2$ model.

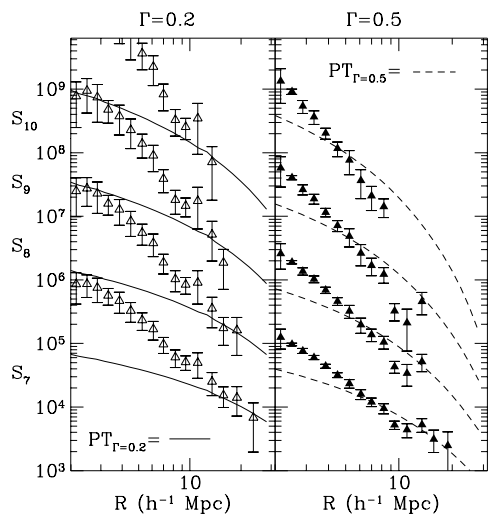

**Figure 5.** Amplitudes $S_J = \overline{\xi}_J / \overline{\xi}_2^{J-1}$ for $J = 7 - 10$ in $\Gamma = 0.5$ (right panel) and $\Gamma = 0.2$ (left panel) N-body simulations.

## 4   DISCUSSION

For both the $\Gamma = 0.2$ and $\Gamma = 0.5$ models, PT gives excellent agreement for $S_{J=3,...,10}$ with the simulations on scales where the variance is approximately linear, i.e. on scales for which $\overline{\xi}_2 \lesssim 1$. More precisely, this scale seems to be $R \simeq 7\,h^{-1}$ Mpc in the $\Gamma = 0.5$ model and $R \simeq 14\,h^{-1}$ Mpc in the $\Gamma = 0.2$ model. For scales $R \gtrsim 30\,h^{-1}$ Mpc, the sampling variance arising from the finite size of our simulation boxes dominates our measurements of the $S_J$. The accuracy of the agreement does not seem to deteriorate too much when the order increases up to $J = 10$.

In Figures 2 and 4 we have compared the results at two different output times, separated by 2.5 expansion factors. The results indicate that the values of $S_J$ do not evolve much with time. This is in spite of the fact that $\overline{\xi}_J$ evolves by a factor $\simeq 10^{J-1}$ between $\sigma_8 = 0.4$ and $\sigma_8 = 1$.

Visually, in Figure 1, the $\Gamma = 0.2$ model *looks* closer to the $\Gamma = 0.5$ than to the clumps model, but nevertheless $\overline{\xi}_2$ for the clumps model is closer to $\overline{\xi}_2$ for the $\Gamma = 0.2$ model, by construction. This illustrates the importance of higher order statistics. It is not difficult to construct artificial point distributions which have similar 2-point statistics to the simulations but very different higher order correlations. For instance, one could place all particles inside a thin wall within the simulated box to produce a given $\overline{\xi}_2$ (see Gaztañaga & Yokoyama 1993). Here we have tried more realistic attempts in order to see if the PT predictions are a general feature of models with "clumps" or a specific feature of gravitational clustering. Figure 3 shows how different the values of $S_3$ for the $\Gamma = 0.2$ gravitational model are compared with the skewness in a clumps models with a similar $\overline{\xi}_2$. We have also generated clumps models with different shapes and amplitudes of $\overline{\xi}_2$ but have not been able to reproduce the values of $S_3$ in either the $\Gamma = 0.2$ or the $\Gamma = 0.5$ models. This is not surprising, as a visual inspection of the clustering pattern in the maps (*cf* Figure 1) shows strong differences between gravitational and clumps models: clusters are not spherical in the N-body simulations.

The values of $S_J$ provide a way to discriminate between different models of structure formation: realistic distributions of points can have significantly (within the sampling errors) different values of $S_J$ at all scales. In particular, we have illustrated here how two distributions with similar values of the 2-point statistics $\overline{\xi}_2$ can produce very different values of $S_J$. The distributions that are evolved under gravity seem to match the PT predictions. In particular, there is a correlation between the shape of $\overline{\xi}_2$ and the values of $S_J$, which we do not find in the mock clumps models. Moreover, these predictions are practically independent of $\Omega$ and $\lambda$ (Bouchet *et al.* 1992, Bernardeau 1994a), in agreement with what we found in the N-body simulations.

The values of $S_J$ in the APM galaxy distribution also seem to follow the PT predictions (Gaztañaga 1994, Gaztañaga & Frieman 1994). To interpret this results one has to consider first the problem of biasing, i.e. how galaxies trace matter. After considering the posibility of both non-linear and non-local biasing, Gaztañaga & Frieman (1994) have found that the simplest interpretation for the values of $S_J$ measured in the APM is that there is little or no biasing on scales $R \gtrsim 7\,h^{-1}$ Mpc, as otherwise one has to invoke a delicate, model-dependent balancing of terms so that the estimated values of $S_J$ follow, at each order, the PT predictions. This interpretation implies that the observed values of $S_J$ in the APM are compatible with the clustering that emerges from gravitational growth of small (initially Gaussian) fluctuations, regardless of the cosmological model we assume for the universe, i.e. $\Omega$, $\lambda$, $H_0$ or the nature of dark matter.

## Acknowledgements

We would like to thank George Efstathiou for supplying us a copy of the $P^3M$ code and Francis Bernardeau for supplying the PT results of $S_J$ for $J > 7$. EG was supported by a Fellowship by the Commission of European Communities. CMB acknowledges the receipt of a PPARC post-doctoral fellowship.